**SUM-BASED SCORING FOR DICHOTOMOUS AND LIKERT-SCALE QUESTIONS**

Tiffany A. Low, Edward D. White, Clay M. Koschnick, and John J. Elshaw

## 1. Introduction

It is a common practice to standardize the inputs of a response to minimize or to remove the influence of differing input scales (Kutner et al., 2004). Typically, this standardization process equates to subtracting the mean from a recorded value and then dividing by the standard deviation (Mendenhall and Sincich, 2007). This term of standardization can also be referred to as normalization, although the meaning of normalization can vary (Dodge, 2003). One common definition of normalization translates that to Min-Max scaling which transforms the recorded input value from the original value to that of a number customarily between 0 and 1 or -1 and 1. One of the main criticisms of either standardization or normalization for a user is that the transformation is akin to a black-box process. Black-box processes are not easily understood or communicated to users and have documented issues (Guidotti et al., 2018).

Likert-scales items are a common measure for elicited responses. To avoid bias, Likert items generally have a balance of both positive and negative responses on a symmetric scale where each successive response option is considered better, although the opposite is true for a reverse Likert scale (Willits et al., 2016). If all Likert items use the same scale, then their responses may be summed together to create a score for each respondent (Johns, 2010; Willits et al., 2016). Statistically, this results in each item have equal variance with respect to the overall variability of the sum. If the Likert-scale items are not of the same scale then unequal variances will occur, indicating that at least one question will have an overly influential effect on the variability of the sum. Such an occurrence could possibility skew any statistical analysis. In this article, we address how to minimize this effect when summing both dichotomous outcomes and a set of similarly scaled Likert responses.

## 2. Methodology

Consider a sum consisting of two sets of independent and identically scaled questions. The first set constitutes Likert-scale questions from 1 to $k$, where $k$ is 3 or greater (integer value). We refer to this henceforth as the ordinal scale of $k$ (taking on possible values of 1, 2, 3, …, $k$). The second set consists strictly of dichotomous values of either 1 or $c$, where $c$ is determined by the user. For example, a respondent has a set of 10 questions: 8 of a Likert-scale ranging from 1 to 5 and 2 from 1 to $c$. Ordinarily, one would intuitively expect a user to set the value of $c$ to $k$. However, the analysis section shows that is almost always incorrect.

To ascertain what value to set $c$ to we adopt the statistical approach of comparing the variance of a single question to the variance of the sum of all questions. Ideally, each question should have a variability weighting of $1/n$ with respect to the overall sum, where $n$ is the number of questions used to arrive at a sum. For simplicity and adoptability, we use ordinal Likert scales with $k$ ranging from 3 – 10. We stop at 10 since we think it would be relatively unrealistic to have a scale greater than that, however, we do present a general formula for going beyond 10 for those interested.

To arrive at the percent contribution for each item, we need to make another assumption besides independence of questions. That assumption assumes that each value of the Likert scale is equally likely compared to another value within that scale. This doesn't imply that a user is equally likely to answer a 1 compared to a *k*. This assumption just implies that the instrument itself does not bias one numerical value over the other. Consequently, this assumption of equally likely implies that the composer of the questionnaire doesn't bias the question prose whereby strongly leading the respondent to one value compared to another based on the composition of the question.

This percent contribution requires the derivation of the mean and variance for each respective Likert scale to determine its actual percent contribution to the response. The mean ($\mu$) or expected value ($E(x)$) of a discrete random variable is

$$\mu = E(x) = \sum xp(x)$$

where *x* is a random variable and *p(x)* is its associated probability (Casella & Berger, 2002; McClave et al., 2018). For the ordinal Likert scale questions *x* varies from 1 to *k*. For the dichotomous, *x* subsumes either 1 or *c*. The probability will be either 1/*k* or ½ since dichotomous questions only have two possible values. The variance of a discrete random variable is defined as the average of the squared distance of *x* from the mean and denoted as:

$$\sigma^2 = E[(x-\mu)^2] = \sum (x-\mu)^2 p(x)$$

Once the variance is calculated for each individual question and then the subsequent variance of the overall sum, the percent contribution of a particular criterion (question) can be found using $R^2$, which represents the coefficient of determination (Kutner et al., 2004). This percent contribution mathematically is a ratio:

$$\% \ Contribution \ = R^2 = \frac{Explained \ Variance}{Total \ Variance} \quad (1)$$

The numerator of (1) does not contain $c$ because the upper value for the dichotomous variable doesn't matter with respect to variance; the range, $r$, representing the difference between 1 and $c$ does. By solving for $r$, we can then back substitute for $c$.

By setting % Contribution to $1/n$, we are then able to solve for $r$, which we define as $c - 1$. Realistically, $r$ will not be an integer. Consequently, we round to the nearest integer. Note: this could involve rounding downward. We recognize that the rounding may not result in optimal mathematical values, however, this action preserves the intent of presenting a non-black-box process when determining the value of $c$ for a user in a questionnaire consisting of a sum of dichotomous inputs and similar scaled Likert questions.

## 3. Results

As noted in Section 2, we do not intermix different ordinal Likert scale questions for it is mathematically impossible for a 3-point, 4-point and 5-point Likert scale (for example) to contribute equally to a sum without using scaling or a variable transformation. As an example, if we want a 4-point Likert criterion to contribute equally to a 5-point Likert criterion, we have to multiply the 4-point Likert criterion by 5 and divide the answer by 4. The more criteria the more complicated this scaling process will be, which makes it less likely for the user to adopt such an approach.

The numerator of (1) mathematically equates to $\frac{r^2}{4}$, where is $r = c - 1$ and $c$ is the upper value of the dichotomous item. The denominator equals $\frac{r^2}{4} n_1 + D n_2$. The number of dichotomous questions is $n_1$, while $n_2$ represents the number of similarly, ordinally scaled Likert questions. Note: $n = n_1 + n_2$; $n$ is the total number of questions constituting the sum. $D$ represents the variance of *k* Likert scale questions; Note: we describe *D* shortly. By setting (1) equal to 1/*n*, we can then solve for *r*. Setting these equal is equivalent to solving $\frac{4D}{r^2} = 1$ with respect to *r*. Because there are two roots to this problem (one negative and the other positive), we elect to present the positive since a range is by definition positive.

For describing *D*, we use symmetry for the variance of an ordinally, scaled Likert question around its mean. We do, however, need to account for when *k* is either odd or even. With that in consideration, *D* is equal to $\sqrt{\frac{8}{k}\sum_{i=1}^{s}\left[i - \left(\frac{k+1}{2}\right)\right]^2}$, where *s* is either (*k*-1)/2 for when *k* is odd and *K*/2 when *k* is even. Table 1 shows the exact values for *c* with the rounded results for users to adopt for *k* = 3 to 10.

**Table 1**: Range, upper value, and suggested value to use for dichotomous questions.

| Scale | Range | Upper Value | Suggested Value |
|---|---|---|---|
| 3 | 1.632993 | 2.632993 | 3 |
| 4 | 2.236068 | 3.236068 | 3 |
| 5 | 2.828427 | 3.828427 | 4 |
| 6 | 3.41565 | 4.41565 | 4 |
| 7 | 4 | 5 | 5 |
| 8 | 4.582576 | 5.582576 | 6 |
| 9 | 5.163978 | 6.163978 | 6 |
| 10 | 5.744563 | 6.744563 | 7 |

**Conclusion**

We investigated whether the values of the dichotomous response mattered if it is desirable for all input variables to have relatively equal contribution to the response variable. We discover that the ranges between the dichotomous values matter more than the values themselves because if two values have the same range, they have the same percent contribution. In addition, the greater the range between the dichotomous variables, the greater its contribution to the response. While it is mathematically feasible to determine a fractional value for the optimal range of a dichotomous variable, a complicated method will unlikely be employed in the field. To simplify the approach, we approximate the optimal range of a dichotomous variable, which depends on the Likert scale used. That in turn generates the suggested upper values of a dichotomous scale, thereby maintaining an equivalency of question contribution when summing dichotomous and non-dichotomous questions in a Likert Scale survey.